\documentclass{PoS}

\title{The X-ray spectro-timing properties of a major radio flare episode in Cygnus X-3}

\ShortTitle{The X-ray spectro-timing properties of a major radio flare episode in Cyg X-3}

\author{\speaker{Karri Koljonen}%
         \thanks{This work is supported by the Finnish Graduate School of Astronomy and Space Sciences.}\\
        Aalto University Mets\"ahovi Observatory, Finland\\
        E-mail: \email{karri.koljonen@gmail.com}}

\author{Michael McCollough\\
        Smithsonian Astrophysical Observatory, USA\\
        E-mail: \email{mmccollough@head.cfa.harvard.edu}}

\author{Diana Hannikainen\\
        Florida Institute of Technology, USA\\
        E-mail: \email{ddcarina@gmail.com}}

\author{Robert Droulans\\
        Lyc\'ee classique d'Echternach, Luxemburg\\
        E-mail: \email{robert.droulans@education.lu}}

\abstract{We have performed a principal component analysis on the X-ray spectra of the microquasar Cygnus X-3 from \textit{RXTE}, \textit{INTEGRAL} and \textit{Swift} during a major flare ejection event in 2006 May-July. The analysis showed that there are two main variability components in play, i.e. two principal components explained almost all the variability in the X-ray lightcurves. According to the spectral shape of these components and spectral fits to the original data, the most probable emission components corresponding to the principal components are inverse-Compton scattering and bremsstrahlung. We find that these components form a double-peaked profile when phase-folded with the peaks occurring in opposite phases. This could be due to an asymmetrical wind around the companion star with which the compact object is interacting.}

\FullConference{ An INTEGRAL view of the high-energy sky (the first 10 years) - 9th INTEGRAL Workshop and celebration of the 10th anniversary of the launch\\
                 15-19 October 2012\\
                 Bibliotheque Nationale de France, Paris, France}

\begin{document}
\sloppy

\section{Introduction}

Modeling the X-ray spectra of X-ray binaries (XRBs) often leads to a problem of degeneracy, i.e. multiple distinct models fit the observed data equally well. A striking example can be seen e.g. in \cite{nowak}, where three very different models are fitted equally well to the same data set of Cygnus X-1, despite the excellent quality of the data that was obtained by all the X-ray satellites in orbit at the time. Similar spectral degeneracy was observed for Cygnus X-3 in \cite{hjalmarsdotter}. Therefore, \textit{even if an apparently good fit is obtained between the data and the model, it does not necessarily imply a match between theory and physical reality.} 

In order to make sense of this degeneracy we need to take other data dimensions into account, namely timing and/or polarization. While we are just reaching the point where the polarization dimension can be explored with very long exposure observations \cite{laurent}, X-ray timing data is readily available, and several methods have been developed to combine spectral and timing analyses (e.g. \cite{vaughan}). These comparisons, however, do not reveal \textit{what the actual spectral components causing the variability are}. 

One way of revealing the variability components of X-ray spectra is to employ principal component analysis (PCA), one of the standard tools of time series analysis that has been introduced in the analysis of the X-ray data of XRBs by \cite{malzac} and references therein, and further refined in \cite{koljonen12} for Cyg X-3 to single out individual emission components causing the variability in the X-ray lightcurves. 

\section{Principal component analysis}

Below, we summarize the main points of the PCA. More detail can be found from \cite{malzac} and \cite{koljonen12}. The first step of the PCA is to arrange a number of spectra measured at different times, i.e. a time series of X-ray spectra, to a data matrix. Then the eigenvectors of the covariance matrix calculated from the original data matrix form the principal components that are responsible for the spectral variation. The accompanying eigenvalue states the proportion of variance of a particular eigenvector, i.e. the highest eigenvalue and accompanying eigenvector is the first principal component of the data set etc. One can then form a linear decomposition of the data set using these eigenvectors, ordered by the proportion of variance. The components producing only a small fraction of the overall variance can then be dropped, reducing the dimensionality, i.e. choosing a small number of eigenvectors for the linear composition. The PCA can be exploited in different ways using both spectral and timing information. In the spectral domain this includes producing the variability spectra and log-eigenvalue diagrams. In the timing domain we can follow the evolution of individual principal components by tracking the value of its normalization.

\section{Applying PCA to Cygnus X-3 data}

\begin{figure}
\begin{center}
\includegraphics[width=0.4\textwidth]{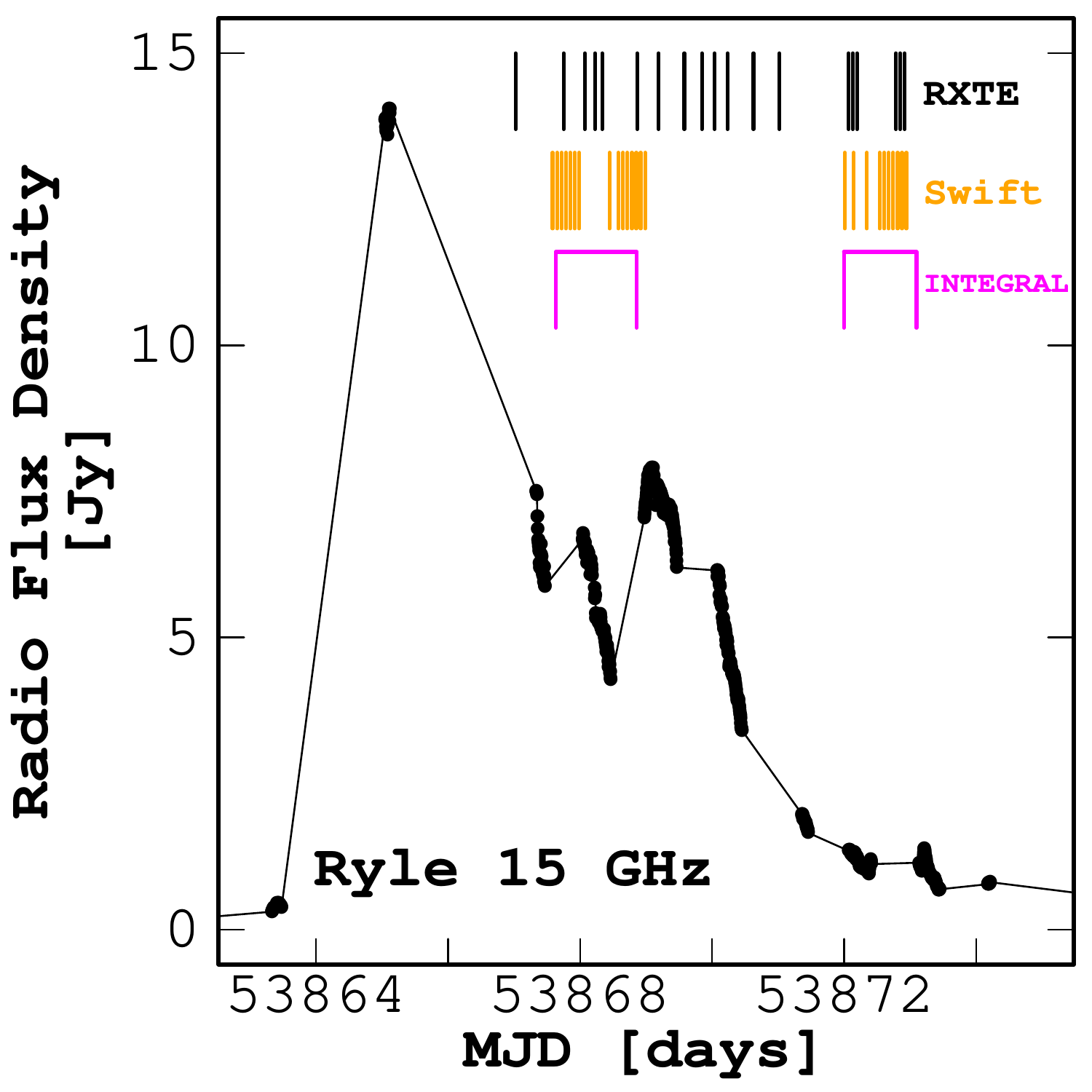}
\end{center}
\caption{The figure show the one-day integrated Ryle/AMI-LA 15~GHz lightcurves from the May 2006 major radio flare. The top part of the panel shows the epochs of the \textit{RXTE} (black), \textit{Swift} (orange) and \textit{INTEGRAL} (magenta) data that are examined here.} \label{obs}
\end{figure}

Cyg X-3 is one of the most peculiar sources amongst microquasars. It is known for massive outbursts that emit radiation from radio to $\gamma$-rays and produce major radio flaring episodes usually with multiple flares that peak up to 20 Jy \cite{waltman}, making it the most radio luminous single object in our Galaxy. The binary components of Cyg X-3 orbit each other in a tight 4.8-hour period \cite{parsignault}, typical for XRBs with a low-mass companion. However, infrared spectral observations suggest that the mass-donating companion in the binary is a massive Wolf-Rayet star \cite{vankerkwijk}. Due to this discrepancy Cyg X-3 is by definition a unique source and similar sources have only been found in other nearby galaxies \cite{cygx3types}. The X-ray spectral and timing properties of Cyg X-3 show a disparity with other XRBs/microquasars increasing the difficulty of interpreting the nature of the system. This disparity could be due to the interaction of the strong stellar wind of the Wolf-Rayet companion with the compact object. 

We have performed a PCA to the X-ray spectra of Cyg X-3 from \textit{RXTE}, \textit{INTEGRAL} and \textit{Swift} during a major flare ejection event. The X-ray observations were obtained during the 2006 May major radio flaring episode, that consisted of two major radio flares with peak flux densities of 13.8~Jy (15~GHz) and 11.2~Jy (11.2~GHz) at MJD 53865 and MJD 53942, respectively (see Fig. \ref{obs} for the May 2006 flare). For \textit{RXTE} and \textit{Swift} data we used the consecutive observations taken right after a flare and for \textit{INTEGRAL} we used revolutions 437, 438 and 462. The radio data are from \cite{koljonen10}.
		  
\subsection{PCA tools of the trade}

\begin{figure}
\begin{center}
\includegraphics[width=1.0\textwidth]{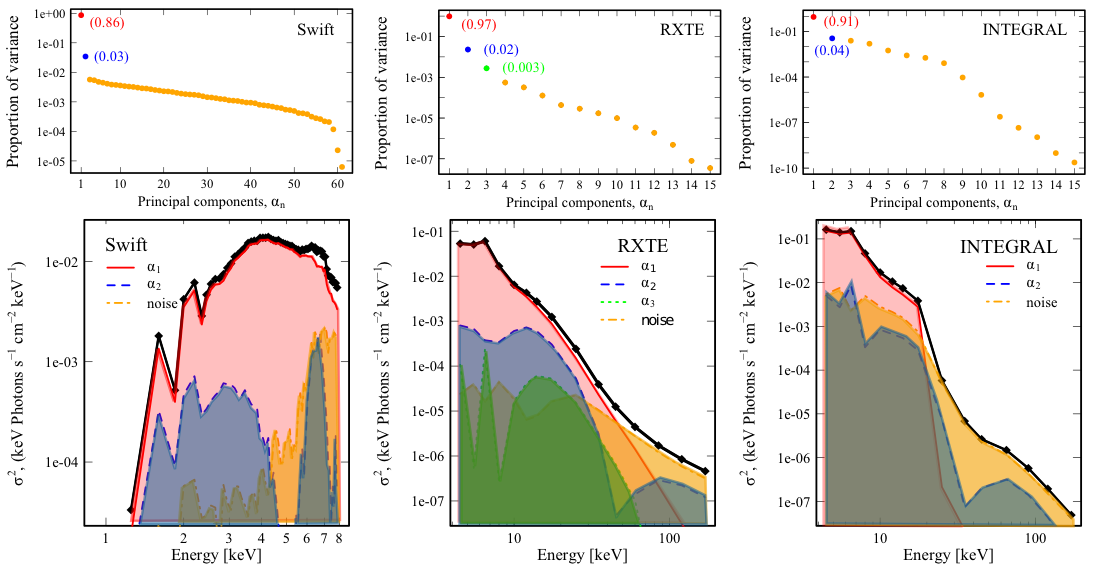}
\end{center}
\caption{\textit{Upper panels:} The LEV diagrams of the principal components from all X-ray observatories (left: \textit{Swift}, center: \textit{RXTE}, right: \textit{INTEGRAL}). The panels show the proportion of variance attributed to each principal component. \textit{Lower panels:} The variance spectra of all observations (black solid line with data points) from all observatories. The colored curves and areas in the figure show the contributions of the principal components $\alpha_{1}$ (red), $\alpha_{2}$ (blue), $\alpha_{3}$ (green) and the remainder of the components totaling to noise and systematic errors (orange).} \label{pcares}
\end{figure}

The standard tool for deciding how many principal components one should retain is called the log-eigenvalue (LEV) diagram (upper panels in Fig. \ref{pcares}). If the principal components decay in geometric progression in the data, the corresponding eigenvalues will appear as a straight line in the LEV diagram and thus signal the start of the ``noise" components. The variance spectrum (lower panels in Fig. \ref{pcares}) is a graph that shows the measured variance as a function of energy. It can be plotted for all the principal components, i.e. showing the overall variance across the energy range of the data, or for each principal component independently thus showing the spectral shape of the varying principal components. From Fig. \ref{pcares} we see that there are two main variability components in play in the Cyg X-3 data, i.e. two principal components explain almost all the variability in the X-ray lightcurves. 

\subsection{The nature of the components}

According to the variance spectral shape of the principal components, a number of models can be ``guessed'' and then fitted to the time-averaged data. The first principal component resembles a Comptonized component and the second a thermal component\footnote{Additionally, a marginal effect ($\sim$1\%) on the variance spectrum is caused by the third principal component in the \textit{RXTE} data which, based on the shape in the variance spectrum, is most likely the reflection component from the accretion disk, although a noise hypothesis is also acceptable.}. Since most of the Cyg X-3 spectra are well fitted by hybrid Comptonization (e.g. \cite{koljonen10, hjalmarsdotter}) we assume that the first principal component is also coupled to this model. However, we need a second spectral component to fit all the spectra successfully and to satisfy the PCA. As the overall effect of the second component in the PCA and X-ray spectra is smaller, multiple components will fit the spectra. We found good fits when using reflection, multicolor disc blackbody, thermal bremsstrahlung, or other thermal Comptonization models in addition to hybrid Comptonization. These best-fitting models and their spectral component normalizations are then compared to the time-averaged evolution of the principal component normalizations. This imposes a second requirement for the X-ray spectral fits, so that in addition to fitting the spectra acceptably, the resulting fits also have to satisfy the spectral evolution inferred from the PCA. This extra requirement reduces greatly, if not completely, the degeneracy of simply using the results from the spectral fits in determining the emission components of the system. Thus, the most probable emission components are those producing the best correlations to the principal component evolution and fits well the X-ray spectra. 

\begin{figure}[h]
 \begin{center}$
 \begin{array}{cc}
 \includegraphics[width=0.5\textwidth]{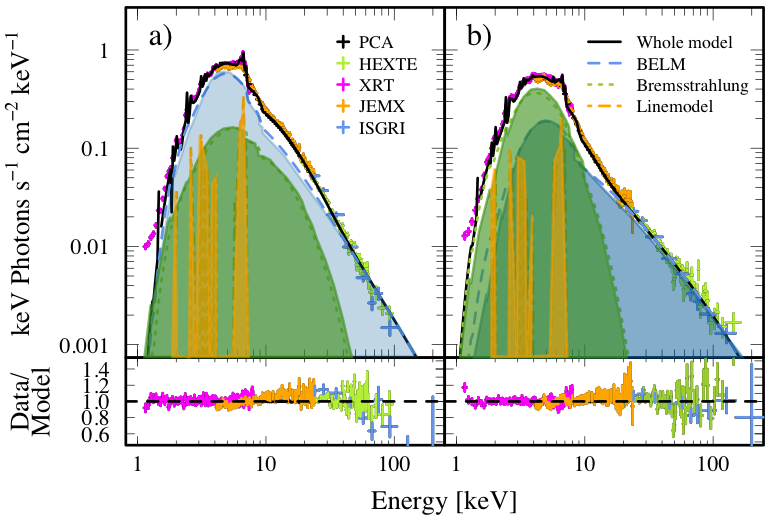} &
 \includegraphics[width=0.35\textwidth]{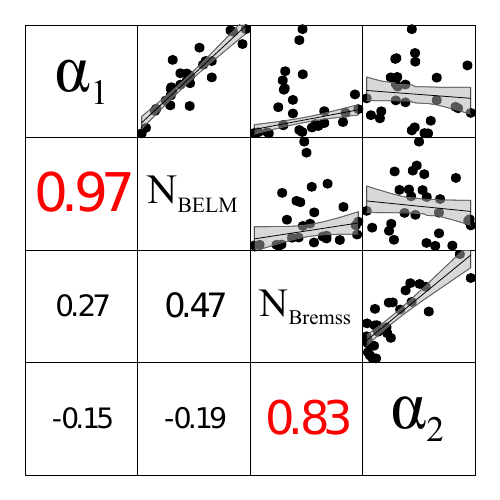}
 \end{array}$
 \end{center}
 \caption{\textit{Left:} An example of two energy spectra with the best-fitting model (black, solid line) and the individual components (Comptonization: blue; bremsstrahlung: green; iron line: orange) overlaid and labelled. \textit{Right:} The scatter plot matrix of the best-fitting model satisfying the principal component evolution. The grid shows the spectral model normalizations N$_{\rm{BELM}}$ and N$_{\rm{Bremss}}$ and the first two principal components ($\alpha_{1}$, $\alpha_{2}$) with robust correlations drawn and written in the appropriate grid cells.} \label{model}
 \end{figure}

To summarize, the best-fitting model in spectral and variability terms has optically thin, rather thermal, Comptonization dominating the variability throughout the X-ray regime and a thermal, rather hot ($\sim$5 keV) plasma component producing variability in the $\sim$10--20 keV regime. We model the hybrid Comptonization component with BELM \cite{belmont} and the thermal component as thermal bremsstrahlung (Fig. \ref{model}). We add to the final model a photoionized emission line model (Savolainen et al., in prep.) and multiply all with a simple absorption model from \cite{koljonen10}.   

\section{Discussion}

\begin{figure}[ht]
 \begin{center}$
 \begin{array}{cc}
 \includegraphics[width=0.28\textwidth]{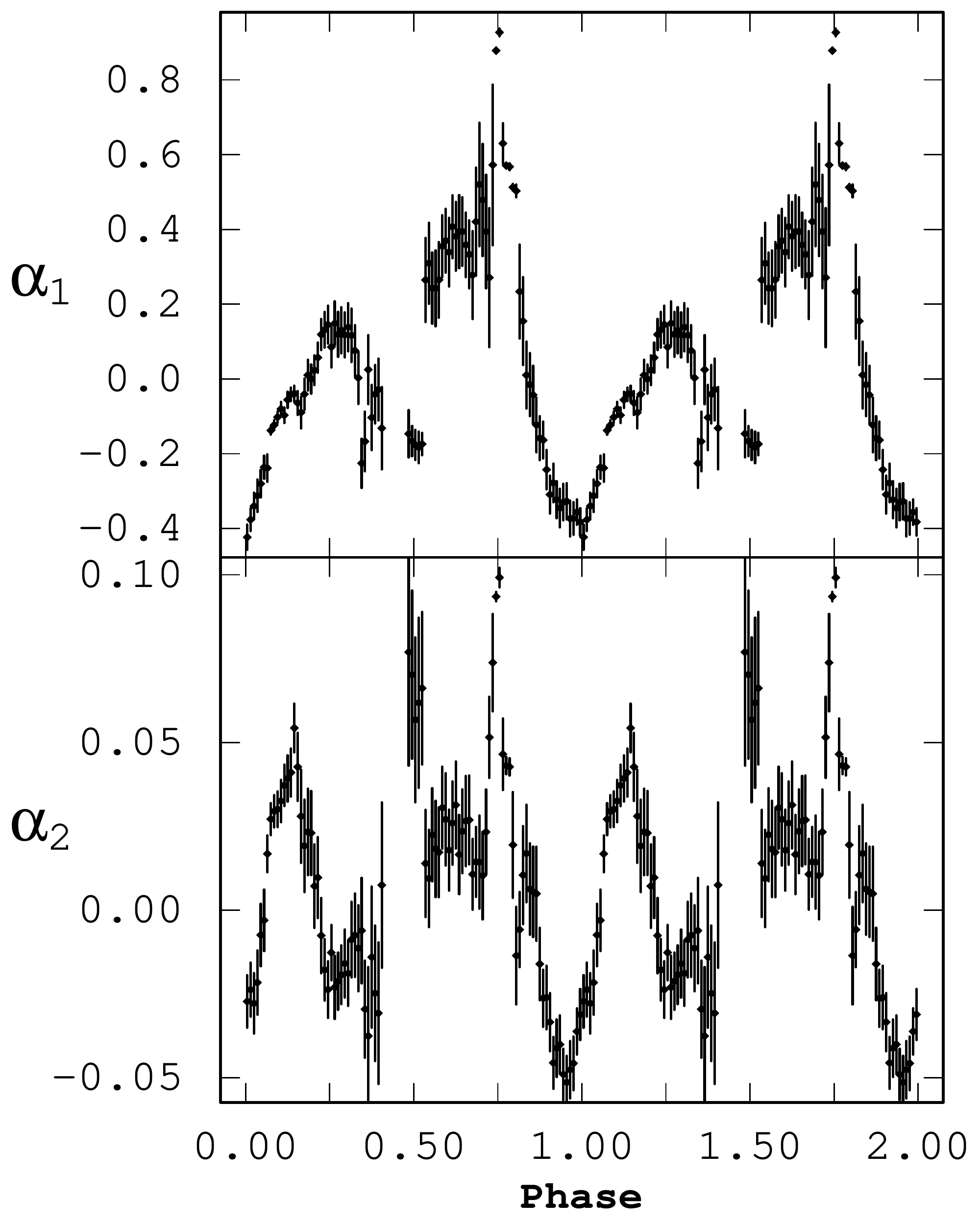} &
 \includegraphics[width=0.5\textwidth]{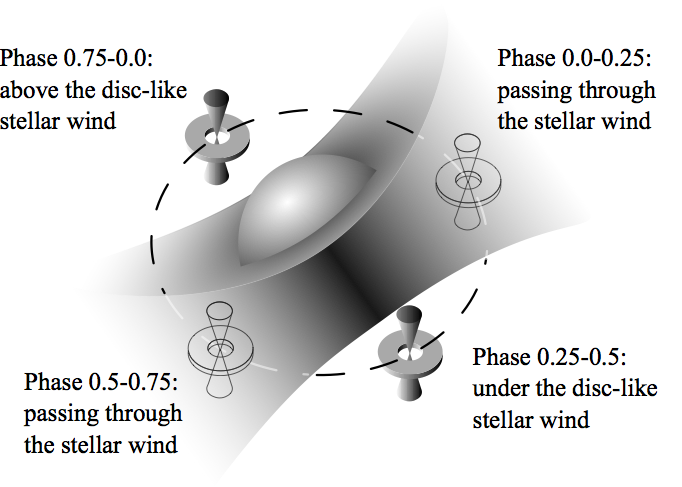}
 \end{array}$
 \end{center}
 \caption{\textit{Left:} The first and second principal component phase-folded through all the \textit{RXTE} data. \textit{Right:} A sketch depicting the geometry of the system, with the Wolf-Rayet companion surrounded by a disc-like stellar wind and the companion object orbiting it (shown in four different orbital phases). Depending on the orbital phase the compact object is either inside (enhanced bremsstrahlung emission) or outside the stellar wind.} \label{phase}
 \end{figure}

The most plausible origin of the thermal component is a plasma cloud that forms as a result of the compact object colliding with the WR stellar wind \cite{zdziarski}. This model was evoked to explain Cyg X-3's lack of high frequencies in the power spectra and the peculiar hard state X-ray spectra with $\sim$ 30 keV cut-off by Compton downscattering. For the plasma parameters found in \cite{zdziarski} the thermal bremsstrahlung emission becomes a substantial source for photons which get upscattered by Comptonization in the plasma cloud.

Due to the short period of the system we can track the principal components through phase. Both components show a double-peaked profile (Fig. \ref{phase}). When relating the second principal component to the bremsstrahlung normalization (see Fig. \ref{model}) which is proportional to $n_{e}n_{i}V$, we see that a change in the density and/or the volume of the bremsstrahlung-emitting plasma is observed along the orbit. The peaks are formed opposite each other, which can be explained by a disk-like shape of the plasma and could arise if the stellar wind is asymmetric.

Similar thermal and hot components have also been found in other microquasars such as GRS 1915$+$105 \cite{titarchuk,mineo}, SS 433 \cite{seifina}, and in several XRBs \cite{zycki}, thus raising questions such as could the thermal, hot component be something intrinsic to microquasars/XRBs. Furthermore, could the emission mechanism be the same and could this scenario be extended to disk winds.

\end{document}